\documentstyle[preprint,aps]{revtex}

\begin{document}
\draft
\preprint{ }
\title{Coulomb Gaps in a Strong Magnetic Field}
\author{S.-R. Eric Yang\cite{byline} and A.H. MacDonald}
\address{Department of Physics, Indiana University, Bloomington, IN
47405}
\date{\today}
\maketitle

\begin{abstract}

We report on a study of interaction effects in the tunneling
density-of-states of a disordered two-dimensional electron
gas in the strong magnetic field limit where only the lowest
Landau level is occupied.
Interactions in the presence of disorder are accounted for
by performing finite-size self-consistent Hartree-Fock calculations.
We find evidence for the formation of a pseudo-gap
with a tunneling density-of-states which vanishes
at the Fermi energy.

\end{abstract}


\narrowtext

In the absence of a magnetic field,
interaction effects in disordered electronic systems
cause dramatic reductions in the tunneling density-of-states (DOS),
$g_T(\epsilon)$, at energies near the chemical potential $\mu$.
In two-dimensions in the weak disorder limit
the relative suppression of $g_T(\mu)$ increases logarithmically
with decreasing temperature\cite{aa,palee}.
At sufficiently low temperatures, where
the weak-disorder approximation fails, all states are localized.
For localization lengths small compared to the distance
between electrons (the classical limit)
the Coulomb gap theory of Efros and
Shklovskii\cite{cgap} is expected\cite{cgapnum} to apply.
In the strong magnetic field limit, the one-body electronic spectrum
in the absence of interactions and disorder
consists of macroscopically degenerate Landau levels
and the behavior of $g_T(\epsilon)$ promises to be even richer.
Weak-disorder theory
gives\cite{girvhough} the same temperature dependence
as at $B=0$, although because of the absence of a small parameter
it cannot be formally justified even
at high temperatures.  At very low temperatures all states
in disordered samples
are expected to be localized except at a single energy within
each Landau level and Polyakov and
Shklovskii\cite{polyshkl} have recently
argued that the low-temperature transport
properties\cite{wei} of disordered samples in strong
magnetic fields provide indirect evidence for Coulomb gap formation.
Direct experimental evidence of strong suppression of
$g_T(\epsilon)$ for $\epsilon$ near $\mu$ comes from
the recent experiments of Ashoori\cite{ashoori} {\it et al.}
and Eisenstein\cite{eisenstein} {\it et al.}
In this Letter we report on the first study
of the suppression of $g_T(\epsilon)$ in a disordered
two dimensional gas of quantum electrons at strong magnetic fields.

We restrict our attention here to the strong magnetic field limit
where the Landau level filling factor, $\nu \equiv N/N_{\phi} < 1$,
the electrons are completely spin-polarized,
and Landau level mixing by interactions or disorder
can be neglected.
(Here $N_{\phi} = B L_x L_y / \Phi_0 \equiv L_x L_y / (2 \pi \ell^2)
$ is the number of
states per Landau level, $\Phi_0 = hc/e$ is the magnetic
flux quantum and $N$ the number of electrons.)
Coulomb interactions are treated self-consistently within
the Hartree-Fock approximation (HFA).  As we discuss
further below the HFA
becomes exact in the classical limit of strongly localized
states and is generally expected to be accurate when the disorder
energy scale is larger than the interaction energy scale.
Calculations have been completed for $\nu =1/5 $ and for
$\nu = 1/2$.  In the former case the states
near the Fermi level in the absence of interactions
have a localization length comparable to the distance between
electrons and we find clear evidence of
the formation of a pseudogap where the DOS
at the Fermi level vanishes in the thermodynamic limit.
As far as we are aware ours are the
first calculations which demonstrate the occurrence of a
Coulomb gap away from the classical limit.
We should expect to find qualitatively similar results
at other values of $\nu$, except possibly at $\nu=1/2$ (see below).

We perform our calculations in the Landau gauge ($\vec A = (0 ,
B x, 0))$ and apply quasiperiodic boundary conditions to
the HF single-particle orbitals
inside a rectangle with sides $L_x$ and $L_y$.  The basis states
used to represent the HF Hamiltonian are related to
elliptic theta functions and can
be labeled by a set of guiding centers $ |X_i\rangle $,
$ i = 1, \cdots , N_{\phi}$ inside
the fundamental cell of the finite-size system\cite{yhl}.
The finite-size system HF calculations in
the absence of disorder have been described in detail
previously\cite{fshf}.  An important simplification related to
the strong magnetic field limit\cite{densmat} is the fact
that both Hartree and Fock potentials can be expressed
as a functional of the electron density:
\begin{equation}
\langle X_j\vert V_{HF}\vert X_i\rangle = \sum_{\vec{q}} \Delta
(\vec{q})\exp{\biggl[\frac{iq_x(X_i + X_j)}{2}\biggr]}\delta (j,i +
m_y)U_{HF}(\vec{q})
\label{eq40}
\end{equation}
Here the sum over $\vec{q}$ is over the discrete set of wavevectors
consistent with the boundary conditions
and $\delta (n',n)$ is $1$ if $n'=n({\rm mod} N_{\phi})$
and $0$ otherwise.  The quantity $\Delta (\vec{q})$ is
proportional\cite{fshf} to the Fourier component of the charge density
at wavevector $\vec q$ and is calculated from
the eigenvectors of the HF Hamiltonian by
\begin{equation}
\Delta (\vec{q})=\frac{1}{N_{\phi}}\sum_{j=1}^{N_{\phi}}
\sum_{j'=1}^{N_{\phi}}\delta (j',j +
m_y)\exp{\biggl[\frac{-iq_x(X_j + X_{j'})}{2}\biggr]}\sum_{\alpha=1}
^{N} \langle
X_{j'}\vert\alpha\rangle\langle\alpha\vert X_j\rangle.
\label{eq50}
\end{equation}
In Eq.~(\ref{eq50}) the sum over $\alpha$ is over the $N$ lowest energy
eigenvectors of the HF Hamiltonian.
$U_{HF}(\vec{q})$ includes both Coulomb and exchange interactions
and needs to be evaluated\cite{fshf} only once for each finite
system size.  $U_{HF}(\vec q)$ is proportional to $e^2/\kappa \ell$
($\kappa$ is the dielectric constant)
which is the characteristic interaction energy scale for interacting
electrons in the lowest Landau level.

We have used a model disorder potential consisting of
$N_I$ $\delta$-function scatterers
with strength uniformly distributed between
$- \lambda$   and $ \lambda $
and scattering centers uniformly distributed
inside the fundamental cell of the finite-size system.
The disorder-averaged DOS in the absence of
interactions is known exactly for this model\cite{brezin}
and similar models have been used previously to study localization
of non-interacting electrons in strong magnetic fields\cite{ando}.
For a given disorder realization the eigenstates
$\vert\alpha\rangle$ and eigenvalues $\varepsilon_\alpha$ of the HFA
satisfy the
following equation:
\begin{equation}
\sum_i \langle X_j\vert H_{HF}\vert X_i\rangle
\langle X_i\vert\alpha\rangle \equiv \sum_i
\biggl[\langle X_j\vert V_I\vert X_i\rangle + \langle X_j\vert V_{HF}\vert
X_i\rangle\biggr]\langle X_i\vert\alpha\rangle = \varepsilon_\alpha\langle
X_j\vert\alpha\rangle.
\label{eq30}
\end{equation}
For each disorder realization Eqs.~(\ref{eq30}), (\ref{eq40}),
and (\ref{eq50}) were solved by iterating to
self-consistently for six different
values $\gamma \equiv (e^2/ \kappa \ell) / \Gamma$ where
$\Gamma = (\lambda^2 N_I
/ \ell^2 L_x L_y)^{1/2}$ is the characteristic disorder potential
energy scale.  (For this model the self-consistent Born approximation
DOS is non-zero for $|\epsilon| < (2 / 3 \pi )^{1/2}
\Gamma \sim 0.46 \Gamma$.)   The HF equations were
solved first at $\gamma = 0$ where no self-consistency is required,
and then at $\gamma = \gamma_i \equiv 0.1 i$, $ i = 1, \cdots, 5$.
For $\gamma= \gamma_i$ the iteration is started using the
self-consistent charge density obtained at $\gamma =\gamma_{i-1}$.
The average change in the charge-density Fourier components at
iteration $j$,
$[\sum_{\vec{q}} \vert\Delta_{j+1}(\vec{q}) -
\Delta_j(\vec{q})\vert] N_{\phi}^{-2} \equiv \delta $,  was evaluated
to test for convergence.  The calculations were considered to
be converged once $\delta < 10^{-6}$.  All the numerical results
reported below were obtained for $N_I=5N_{\phi}$.  Square ($L_x=L_y=L$)
systems of various sizes were considered for both $\nu=1/5$ and
$\nu = 1/2$.

For $\nu$ or $1 - \nu$ very small the classical limit is approached.
In this limit the HF energy for the
orbital at disorder site $i$ is
\begin{equation}
E_i=\phi_i + \sum_j [\langle ij\vert V\vert ij\rangle - \sum_j \langle
ij\vert V\vert ji\rangle] f_j
\label{eq90}
\end{equation}
where $\phi_i$ and $f_i$ are the disorder-potential
energy at site $i$ and the occupation number at site $i$.
If the overlap between different orbitals can be neglected
the exchange matrix element in Eq.~(\ref{eq90}) is zero
unless $i=j$ and we obtain
$E_i=\phi_i + \sum_{j\neq i} V_{ij}f_j$ where $V_{ij}=\langle
ij\vert
V\vert ij\rangle$.  In the localized limit the
only effect of exchange is to cancel the self-interaction
energy of each orbital.  This is just the classical expression
for the energy of an electron at site $i$ used in previous
investigations of the Efros-Shklovskii Coulomb gap\cite{cgapnum}.
Thus the  HF approximation captures all the
physics of the classical limit but is able to include quantum effects
without additional difficulty.

Typical numerical results are shown in Fig.~(\ref{fig1})
for $\nu = 1/5$ and in Fig.~(\ref{fig2}) for $\nu = 1/2$.
In the HFA the tunneling
DOS is given by the spectral density of HF
eigenvalues.  We plot the total number of eigenvalues which occur
in a range of width $\Delta E$ around the energy $E$,
$D_{\Delta E}(E)$, for all disorder realization studied.
(Note that the histogram
box widths used to obtain these results are several times larger
than the typical energy level spacing at the Fermi energy.)
For a given system size the DOS
exhibits a minimum which is tied to the Fermi level and which
deepens and broadens as the interaction strength increases or as the
disorder decreases.  For a given ratio of
interaction strength to disorder the DOS
minimum becomes more pronounced as the system size increases.

In the classical limit the difference between the energy of
an orbital which is occupied and an orbital which is empty must
exceed\cite{cgap,cgapnum} the excitonic energy associated
with the transfer of an electron from the occupied site
to the empty site.  For
a finite size square system with periodic boundary conditions
the minimum possible excitonic energy (and hence the
minimum possible level spacing) occurs when the displacement
of the charge is $(L,L)/\sqrt{2}$ and equals
$ \sim 2.285 e^2 / \kappa L$.
This energy provides a lower bound on the energy-level-spacing
at the Fermi level in a disordered system of classical
electrons.
In the thermodynamic limit Efros and Shklovskii have estimated
that for classical electrons the impurity-averaged
DOS at the Fermi level
in two-dimensions vanishes linearly with a coefficient
which corresponds to an energy level spacing at the Fermi level
of $ \sqrt{2 \pi} e^2 / \kappa L \sim 2.507 e^2 / \kappa L$.
The dependence of our HF spectrum DOS
minimum on system size is illustrated
in Fig.~(\ref{fig3}) where we plot the DOS
per Landau Level at the Fermi energy
as a function of $x \equiv (\Gamma /N_{\phi}) /
(e^2 /  \kappa L )$.
$x$ is proportional to the ratio of the energy-level-spacing
in a finite system of electrons without interactions to the
energy-level-spacing at the Fermi level for a disordered
system of classical electrons with Coulomb interactions.
We see from Fig.~(\ref{fig3})
that to a good approximation the Fermi level DOS
depends only on this ratio.  For large $x$ Coulomb interactions
are unimportant and $g_T(\epsilon_F)$ approaches the known
result for non-interacting electrons, even for quite small
system sizes.  For small $x$ the numerical results suggest
that $g_T(\epsilon_F)$ is proportional to $x$ with a slope
corresponding to an energy-level spacing at the Fermi level
of $ \sim 0.8 e^2 / \kappa L$ for $\nu = 1/5$ and
$ \sim 0.6 e^2 / \kappa L$ for $\nu =1/2$.
This decrease in coefficient of the $L^{-1}$
energy-level-spacing at the Fermi level
for large $L$ compared to the classical limit
cannot be explained by the finite spatial extent
of the quantum electronic orbitals alone\cite{cgapnum}
and is a non-trivial quantum effect.  However the $L^{-1}$
dependence of the energy-level spacing on system size
is not altered by quantum effects; we believe that these
numerical results convincingly establish that the
HF DOS vanishes at the Fermi level
in the thermodynamic limit\cite{further}.

For non-interacting electrons in the lowest Landau level
states at the Fermi level are localized except
(for typical disorder potentials) at $\nu =1/2$.
It is therefore surprising that our numerical results at
$\nu=1/5$ and $\nu=1/2$ are not more different.  For
$\nu=1/2$ the conductivity is finite\cite{huobhatt} and
the two-particle spectral response has a diffusive form
at long distances\cite{chalker}.  To lowest order in interactions
we can construct the HF self-energy from the
non-interacting system eigenstates.  Following work on interaction
effects in disordered metals this leads\cite{palee} to
a DOS change from the exchange self-energy
\begin{equation}
\frac{\delta D(E)}{D(E)}\sim\frac{-\partial\sum_{X} (E)}{\partial E}
\sim\frac{e^2}{\hbar D}\int_{q_c}
\frac{dq}{q^2}= \frac{e^2/\kappa \ell}
{( \Gamma |E - \epsilon_F|)^{1/2}}.
\label{eq80}
\end{equation}
Eq.~(\ref{eq80}) is a good agreement with our numerical results
at $\nu = 1/2$ far enough from the Fermi energy where it
is permissible to treat interactions in lowest order.
In Eq.~(\ref{eq80}) we have evaluated the exchange self-energy
using the bare two-dimensional $1/r$ interaction, $V(q)= 2 \pi e^2 /q$.
It is the absence of screening in the HFA
which makes the DOS change increase more strongly
as energies approach $\epsilon_F$ than the $ \ln (E)$ dependence
expected for disordered metallic systems in two-dimensions.
We expect that screening effects neglected in the HFA
would increase the static dielectric
constant, $\kappa$, and change our results quantitatively
but not qualitatively for $\nu \ne 1/2$.
For $\nu =1/2$, however, the localization length diverges and
$\kappa$ may also diverge\cite{screen}. Further work
will be required to determine whether $g_T(\epsilon)$
at $\nu \equiv 1/2$ is anomalous.

Finally we mention the downward shift in the chemical potential
due to interactions which is visible in Fig.~(\ref{fig1}) and
in Fig.~(\ref{fig2}).  For the case of $\nu = 1/5 $
the magnitude of this exchange self-energy shift
in the chemical potential is much smaller than the value
$ \nu \sqrt{\pi/2} (e^2 / \kappa \ell)$ calculated
neglecting spatial correlations of the occupied orbitals and
used in many considerations of the competition between
disorder and weak interactions.  This reduction has
important implications for the compressibility of
disordered electronic systems in a strong magnetic field
and for the magnitude of exchange-enhanced spin-polarizations.

In conclusion we have performed finite system self-consistent
HF calculations for a disordered system of
two-dimensional quantum electrons in the strong magnetic field
limit.  We find that the HF energy-level
spacing at the Fermi level varies as $e^2 / \kappa L$,
corresponding in the thermodynamic limit to a tunneling
DOS per area which vanishes linearly as the
Fermi energy is approached, for all filling factors.
We expect that these results apply to systems with
sufficient disorder that the fractional quantum Hall effect
does not occur.  This reduction in $g_T(\epsilon)$
near the Fermi energy results from static correlations in
the self-consistent HF ground state which limit
the phase space for adding an electron near the chemical
potential.  Similar reductions in
$g_T(\epsilon)$ should occur for any correlated electron state,
although the behavior of $g_T(\epsilon)$ at very low energies
could differ qualitatively in the strongly correlated fluid
states which occur in weak-disorder systems.

The authors acknowledge stimulating and instructive interactions
with B.L. Altschuler, R.C. Ashoori, A.L. Efros, J.P. Eisenstein,
and B.I. Shklovskii.
This work was supported by the National Science Foundation
under Grant No. DMR-9113911.

\newpage

\begin{figure}
\caption{Tunneling density-of-states histograms obtained from
self-consistent Hartree-Fock calculations at $\nu=1/5$.
$D_{\Delta E} (E)$ is the number of Hartree-Fock eigenvalues
at energies between $E$ and $E + \Delta E$ including all
disorder realizations.  These results are for $N_{\phi}= 245$
and $N_D = 90$ disorder realizations.  For these plots
$\Delta E =0.028 \Gamma$.}
\label{fig1}
\end{figure}

\begin{figure}
\caption{Tunneling density-of-states histograms for $\nu = 1/2$,
$N_{\phi} = 288$, $N_D=60$, and $\Delta E = 028
 \Gamma$. }
\label{fig2}
\end{figure}

\begin{figure}
\caption{Finite system density-of-states minima versus $x$.
At a given filling factor results at different values of
$\gamma$ and different system sizes collapse onto a single curve.
$D_{\Delta E} (E_F) \Gamma / \Delta E N_{\phi} N_D $ is the density
of states per flux quantum per $\Gamma$ energy interval.
The horizontal arrows at the right hand side of the figure
indicate the exact values for the large $x$ (no interaction),
large-system limit.  The plotting symbols are horizontal bars
($\gamma =0.1$), crosses ($\gamma = 0.2 $), diamonds ($\gamma = 0.3$),
squares ($\gamma =0.4$), and octagons ($\gamma =0.5$).
The solid lines are guides to the eye. For $\nu =1/2$ the plotted
$(N_{\phi},N_D)$ are (128,90), (288,60), and (450,7).
For $\nu = 1/5$ they are (80,150), (125,120), and (245,90).
($ x = \gamma^{-1} (2 \pi / N_\phi)^{1/2} $.) The dashed line
is the Efros-Shklovskii expression for classical electrons.}
\label{fig3}
\end{figure}


\begin{references}
\bibitem{further} Further work aimed at studying quantitatively
the dependence on filling factor and its relation to the
diverging localization length at the Landau level center is
in progress.
\end{references}
\end{document}